\documentstyle[12pt]{article}
\begin{document}

%begin macros
\def\nn{\nonumber \\}
\def\be{\begin{equation}}
\def\ee{\end{equation}}

\def\bd{\begin{displaymath}}
\def\ed{\end{displaymath}}
\def\ba{\begin{eqnarray}}
\def\ea{\end{eqnarray}}
\def\la{\label}
\def\re{(\ref}

\def\i{{\rm i}}
\let\a=\alpha \let\b=\beta \let\g=\gamma \let\d=\delta
\let\e=\varepsilon \let\z=\zeta \let\h=\eta \let\th=\theta
\let\dh=\vartheta \let\k=\kappa \let\l=\lambda \let\m=\mu
\let\n=\nu \let\x=\xi \let\p=\pi \let\r=\rho \let\s=\sigma
\let\t=\tau \let\o=\omega \let\c=\chi \let\ps=\psi
\let\ph=\varphi \let\Ph=\phi \let\PH=\Phi \let\Ps=\Psi
\let\O=\Omega \let\S=\Sigma \let\P=\Pi \let\Th=\Theta
\let\L=\Lambda \let\G=\Gamma \let\D=\Delta

\def\0{\over } \def\1{\vec } \def\2{{1\over2}} \def\4{{1\over4}}
\def\5{\bar } %\def\5{\overline }
\def\6{\partial }
\def\7#1{{#1}\llap{/}}
\def\8#1{{\textstyle{#1}}} \def\9#1{{\bf {#1}}}

\def\({\left(} \def\){\right)} \def\<{\langle } \def\>{\rangle }
\def\[{\left[} \def\]{\right]} \def\lb{\left\{} \def\rb{\right\}}
\let\lra=\leftrightarrow \let\LRA=\Leftrightarrow
\let\Ra=\Rightarrow \let\ra=\rightarrow
\def\ul{\underline}

\let\ap=\approx \let\eq=\equiv  %% \let\ex=\times \let\hc=\dagger
        \let\ti=\tilde \let\bl=\biggl \let\br=\biggr
\let\bi=\choose \let\at=\atop \let\mat=\pmatrix
\def\CL{{\cal L}} \def\CD{{\cal D}} \def\rd{{\rm d}} \def\rD{{\rm D}}
\def\CH{{\cal H}} \def\CT{{\cal T}} \def\CC{{\cal C}} \def\CM{{\cal M}}
%end macros

\begin{titlepage}
\renewcommand{\thefootnote}{\fnsymbol{footnote}}
\renewcommand{\baselinestretch}{1.3}
\hfill  TUW - 93 - 08 \\
\medskip
\hfill hepth@xxx/9307025 \\
\medskip
\vfill

\begin{center}
{\LARGE {Euclidean 2D Gravity with Torsion}}
\medskip
\vfill

\renewcommand{\baselinestretch}{1} {\large {PETER
SCHALLER\footnote{e-mail: schaller@email.tuwien.ac.at} \\
\medskip THOMAS STROBL\footnote{e-mail:
tstrobl@email.tuwien.ac.at} \\ \medskip\medskip
\medskip \medskip
Institut f\"ur Theoretische Physik \\
Technische Universit\"at Wien\\
Wiedner Hauptstr. 8-10, A-1040 Vienna\\
Austria\\} }
\end{center}

\vfill
\begin{center}
submitted to: {\em Phys. Lett.}
\end{center}
\vfill
\renewcommand{\baselinestretch}{1}                          %!!!

\begin{abstract}
Closing a gap in the literature on the subject, the local solutions of
2D-gravity with torsion are given for Euclidean signature. For the
topology of a cylinder the system is quantized in a Dirac approach.
\end{abstract}

\vfill
\hfill Vienna, April 1993  \\
\end{titlepage}

\renewcommand{\baselinestretch}{1}
\setcounter{footnote}{0}

The main reason to consider  \cite{Vol}
\be
{\cal L} \, =  e \, (-\frac{\gamma}{4} \, R^2 +
\frac{\beta}{2} \,T^2 - \lambda)
 \la{action}  \ee
was to provide a 'kinetic term' for the gravitational sector of
2D string theory. Whereas, however, for Lorentz signature of the world
sheet metric $g_{\m\n}$ {\em all} the classical and quantum solutions
to \re{action}) have been found (cf.  \cite{Kat}, \cite{Kum},
\cite{Nov}, \cite{All},
\cite{Jap}, \cite{Can}, \cite{Fun}, and references
therein), the corresponding solutions
for Euclidean signature have been still missing up to now. It is the
incentive of this letter to fill this gap.

The basic quantities in \re{action}) are
the orthonormal one--form $e^a$ and  and the SO(2) connection $\o$;
from them one obtaines  the Ricci
scalar $R = 2 \ast {\rm d} \o$,
the 'torsion function' $T^a = \ast {\rm D} e^a$, and $e \equiv \det
e_\mu{}^a$.\footnote{In \cite{Vol} the second term of \re{action})
is rather $-e{\b \0 4}T^{a\m\n}T_{a\m\n} = -e(-)^M \frac{\beta}{2} \,T^2$
with $\det g_{ab} = (-)^M$. We changed the sign of $\b$ in the Euclidean
case $M=0$, since only then the {\em Hamiltonian} formulations differ
just by the different tangent space metric $g_{ab}=diag(1,(-)^M)$ and can be
compared more easily.}
The first order  form of \re{action}) is
\be
\CL_H = e({\pi_2 \0 2} R + \pi_a T^a + E), \la{LH} \ee
with
\be  E \; \equiv \; \frac{1}{4\gamma} \, (\pi_2)^2
- \frac{1}{2\b} \, \pi^2 - \lambda,
 \la{E}
\ee
from which one infers the following canonical structure:
The phase space is spanned by the canonically conjugates
$(e_1{}^a,\o_1;\pi_a,
\pi_2)$  subject to the first class constraints
\addtocounter{equation}{1}
$$ G_a \,=\,  \partial \pi_a +
\varepsilon_{ab} \,  E \, e_1{}^b + \varepsilon_{ab}  \, \pi_b \,
\omega_1  \; \approx \; 0 \eqno(\arabic{equation}a) $$
$$ G_2 \,=\, \partial \, \pi_2 +
\varepsilon_{ab} \, \pi_a \, e_1{}^b
\; \approx \; 0. \eqno(\arabic{equation}b) $$
The Hamiltonian is
\be \CH = - e_0{}^a G_a - \o_0 G_2 + \6 (e_0{}^a \pi_a + \o_0 \pi_2)
\la{H}, \ee
and $e_0{}^a$ and $\o_0$ play the role of Lagrange multipliers.

To have a well-behaved Hamiltonian formulation (ensuring e.\ g.\ that the
reduced phase space is even dimensional, a necessary condition
for a symlectic structure on it) one needs also specific
boundary conditions.  One possibility is to compactify the
$x^1$-direction (topology of a cylinder), which is implemented
by means of periodic boundary conditions in the $x^1$-variable. This option
will be chosen when quantizing the model.
In the first part of this letter, however, we will be interested only
in {\em local} classical Euclidean solutions of \re{action}), postponing
a more elaborate global analysis to some later work. The boundaries
need less careful treatment then and the gauge freedom present locally
(i.e. in {\em some} surrounding of each point) is usually
larger than  the one present globally, since in the latter case
some specific  boundary conditions must not be violated by the gauge
transformations.

To better understand this last point as well as to
get some feeling for the 'physical' (i.\ e.\ gauge independent)
content of the theory, let us have recourse to the canonical abelianization
of the Minkowskian version of the model as found in \cite{Can}.
Introducing
\be \pi_\pm = {1\0 \sqrt{2}} (\pi_0 \pm i \pi_1) \la{pm} \ee
 etc.\, we can, on all of the (complexified) phase space except
 for points with $\pi^2 \neq 0$, use either of the sets of new
 {\em canonical} variables
\be ({\pm iG_\pm\0\pi_\pm}, \mp i{G_2\0\pi_\pm}, Q; \pi_2,\pi_\pm,P^{(\pm)}),
\la{can} \ee
with\footnote{The normalization of $Q$ is  different
to the one in \cite{All}, \cite{Can}, where it was chosen so
as to coincide with the constant $A$ in Katanaev's work. Here we
multiplied it by $(\b^3/4\g)$ so that it is better behaved under
contractions of \re{action}) to $R^2$-Gravity, the Jackiw--Teitelboim model,
and other models, as  reported elsewhere \cite{Lim}.}
\be Q = -\b \exp({\pi_2\0\b})[E - {\b\0 2\g}\pi_2 + {\b^2 \0 2\g}], \qquad
 P^{(\pm)}= - \exp(-{\pi_2\0 \b}) \, {e_1{}^\mp \over \pi_\pm}. \la{Q} \ee
The factors $i$ in \re{can}) come about since here
 $\e_{+-}=-i$ when  $\e_{01} =1$.
Since
\be  \6 Q =\exp({\pi_2\0\b})[\pi^a G_a - E G_2]  \approx 0, \la{dQ} \ee
all of the new generalized coordinates are (strongly commuting)
first class constraints except for the constant part $Q_0$ of $Q(x^1)$,
which  is   a Dirac observable. With regard to the gauge independent
portion of the corresponding momenta, however, the above mentioned
dependence on the boundaries appears: Whereas locally $\6 Q$ can shift
{\em all} of $P^{(\pm)}$ [locally any (non-diverging) function  has
an integral], periodic boundary conditions obviously make the
(otherwise ill-defined) zero mode $P^{(\pm)}_0$ insensitive to
gauge transformations.

Let us dwell
some more on the latter case of a compact $x^1$-direction. Due to the absence
of surface contributions as
well as $Q(x^1) \approx Q_0 = const$ one verifies
 $P^{(+)}_0 \approx P^{(-)}_0$, so that we can also use
\be  P_0 := \2 (P^{(+)}_0 + P^{(-)}_0) = -\oint_{S^1}
 \exp(-{\pi_2\0 \b}) \, {\pi_ae_1{}^a \over \pi^2} \ dx^1 \la{P0} \ee
as the momentum conjugate to $Q_0$. $P_0$ has the advantage of being
already real such as $Q_0$. Thus it is these two coordinates spanning
the reduced phase space (RPS) of Euclidean 2D gravity with torsion on the
cylinder, although only locally (in the RPS) since we operated out
points with $\pi^2 =0$ to obtain the  simple  canonical
splitting in \re{can}).
To interpret $P_0$ let us focus on such global solutions which
allow for closed curves $\CC$ of constant curvature $\pi_2$
(which is at least the case when $\pi^2 \neq 0$ everywhere on
the space--time manifold, as becomes clear from the solutions
below): Then one can check that  $P_0\!\mid_{\CC} \: =
\oint_{\CC} \sqrt{g_{11}} dx^1$.
That is to say, the
quantity $P_0$ is a measure for the  metric
determined circumference of the cylinder; as such it is a gauge independent
quantity, but intrinsically global.

We could proceed using some
compexification argument to obtain the classical Euclidean solutions
from the already known Minkowskian ones \cite{Kat}, \cite{Kum}.
But this shell be shall be postponed to an appendix for the following
reason:
Within the conformal gauge \cite{Kat}, for which a 'Wick rotation' poses no
problem,
the Minkowskian solutions have been found only up to a first
order differential equation. Moreover, lines with $\pi^2=0$ could be covered
only by cumbersome glueing
of the local solutions (valid for $\pi^2 \neq 0$) via geodetic completing
\cite{Com}; the
$C^\infty$--structure of the resulting space--time manifolds $\CM$
remained unproven in this way. Within the 'light cone gauge'
$e_0{}^+=1,\, e_0{}^-=\o_0=0$
\cite{Kum}, on the other hand, a simple algebraic solution was available for
$\pi_+ \neq 0$, but also in the vicinity of points with $\pi_a =0$ \cite{Can}.
Solutions such as the latter allowed to complete the proof of the
$C^\infty$--structure of $\CM$ \cite{Klo}.
In the  case of the light cone gauge, however, a continuation to the Euclidean
solutions is more complicated as one might think at first sight
(cf.\ Appendix). Since a direct  Hamiltonian calculation of
the Euclidean solutions is straightforward and yields also  algebraic
solutions, we will present this possibility in the following.

To find the classical solutions within the Hamiltonian framework
there are basically two (equivalent) procedures: The first more
traditional one, which is also
closer related to the Lagrangian formulation, fixes the Hamiltonian
by means of a specific choice for the Lagrange multipliers. The flow of $\CH$
on the constraint surface yields then the $x^0$-dependence of the basic
fields, and
the still remaining   gauge equivalence of the 'initial data' is factored
out in a final step. The alternative procedure, the advantages of which
we pointed out already in \cite{Can}, starts from a
gauge fixation of the constraints within the phase space; identifying then
a gauge fixing parameter with the evolution parameter $x^0$ allows
[for the case of independent (effective) constraints]
to determine the Lagrange multipliers by  purely  algebraic means.
Since the constraints are not completely independent in our
case\footnote{Integrating \re{dQ}) over $S^1$ reveals that
 the constraints are dependent globally; dropping
possible but ill-defined boundary contributions when
calculating the flow of the Hamiltonian (there appear $\6 \d$--terms
to be integrated by parts)  this dependence becomes visible
also on a local level.}
there remains one ordinary differential equation in $x^1$
for the Lagrange multipliers
yielding an integration 'constant' $f(x^0)$, which has to be gauge fixed
as well. As a last step one then has  to integrate the flow of $\CH$
for the remaining  phase space variables.
It is the second method which turns out to be so powerful in our context.

First we have to find good gauge conditions. In a region where the curvature
is non-constant and  has also no extremal point, slices of constant $\pi_2$
provide a good (local) foliation. (Note that according to \re{LH}),\ \re{E})
on--shell $\pi_2=-\g R$,  $\pi_a=\b T_a$).
Further, by SO(2) rotations of the zweibein frame we  can {\em always} manage
to set $\pi_0 = 0$; and for the case that $\pi_1 \neq 0$
this can be made a perfect cross section to the frame rotations, when
eliminating the Gribov ambiguity $\pi_1 \to -\pi_1$ by the
requirement $\pi_1 >0$. The remaining gauge freedom $x^1 \to a(x^\m)$ can
be used to obtain $e_1{}^1 =1$,  since the differential equation
$e_1{}^1(x^0, a(x^\m)) \6_1 a(x^\m) = 1$ is known to
always have a solution for $a$ locally. For the case of a
closed $x^1$ direction, however, it is at this point where one
could obtain only $\6e_1{}^1=0$, since the zero mode $\oint
e_1{}^1 dx^1$ is diffeomorphism invariant. So let us start with
\be\pi_0 = 0, \qquad  \pi_2 = x^0, \qquad e_1{}^1 =1. \la{gauge} \ee
Note that there still remains the gauge freedom $x^1 \to x^1 + F(x^0)$, which
we will make use of below when determining the Lagrange multipliers.

{}From the 'Faddeev-Popov matrix'
\be \left( \matrix{\{\pi_2,G_0 \}& \{e_1{}^1,G_0 \}&\{\pi_0,G_0 \} \cr
                  \{\pi_2,G_1 \}& \{e_1{}^1,G_1 \}&\{\pi_0,G_1 \} \cr
                  \{\pi_2,G_2 \}& \{e_1{}^1,G_2 \}&\{\pi_0,G_2 \} \cr}
                  \right) =
\left( \matrix{ -\pi_1 \d& 0&0 \cr
                  *&-\6 \d&0 \cr
                  0&*&\pi_1 \d \cr}                   \right) ,
\la{fp}
\ee
in which we used the constraints as well as the gauge conditions
\re{gauge}), we learn
that  our gauge breaks down {\em only} when $\pi_1$ becomes
zero, i.\ e.\ on-shell there is no need for further restrictions on the
curvature.\footnote{At first sight one could think $\det$\re{fp})$=0$
since $Ker \, \6 \d \neq 0$ on $C^\infty(R)$ and thus $\6 \d$
is not
invertible. Note, however, that $Im \, \6 \d=C^\infty(R)$. This is
the relevant condition for the attainability of the gauge. So, calculating
the Faddeev-Popov determinant, we should regard $\6 \d$ as an operator
$\6 \d:C^\infty(R)/Ker \, \6 \d  \to C^\infty(R)$ rather than as an operator
$\6 \d:C^\infty(R)  \to C^\infty(R)$. This subtlety arises, as we did not
specify boundary conditions.}
Indeed, a (gauge independent) analysis of the constraints and
the flow of the Hamiltonian, or, equivalently, the covariant form
of the field equations (cf.\ e.\ g.\ \cite{All})
\be \pi_{a;b} =- \e_{ab} E , \qquad \pi_{2,a} = \e_{ab} \pi^b,  \la{cov} \ee
reveal that points with $\6_\m \pi_2 =0$
are always points with vanishing torsion $\pi_a =0$.
Furthermore, all solutions to the field equations \re{cov}) with  constant
curvature are   $\pi_a \equiv 0, \pi_2  \equiv \pm \sqrt{4\g\l}$. These
describe a general two dimensional Euclidean (Anti-)de-Sitter  space.
Being well-known already, we will not treat these solutions here further.
The analysis of the solutions around extremal points of $\pi_2$,
on the other hand, shall be taken up after exploring the gauge
\re{gauge}).

Inverting $G_0 =0$ and $G_2=0$ in the above  gauge \re{gauge}), one obtains
\be \o_1 = -{E \0 \pi_1} , \qquad e_1{}^0 =0,   \la{invert} \ee
respectively. Due to \re{dQ}) we can  use the first
equation of  \re{Q}) instead of $G_1=0$, yielding
\be \pi_1 =   \sqrt{W(x^0)}, \qquad W(x^0) \equiv
2Q_0 \exp({-x^0 \0 \b}) + {\b \0 2\g}(x^0)^2 -
{\b^2 \0 \g}(x^0-\b) -2\b\l, \la{pi1} \ee
in which $Q_0$ plays the role of an integration constant
(due to \re{H}) $Q_0$, being a Dirac observable, is also a constant of the
motion).
The 'Lagrange multipliers' are now obtained by differentiating
\re{gauge}) with respect to $x^0$ and requiring that $\6_0$ is
generated by the Hamiltonian. This yields two algebraic and one
ordinary differential equation in $x^1$ for the Lagrange
multipliers, the latter giving rise to an undetermined function
$f(x^0)$:
\be e_0{}^0 = {1\0\pi_1}  , \qquad e_0{}^1 =-({E\0(\pi_1)^2} + {1\0 \b})x^1
+f(x^0) , \qquad
\o_0 = - {e_0{}^1 E\0\pi_1} \la{lm}. \ee
It is straightforward to check that $f$ can be made to vanish
by means of the
residual gauge freedom $x^1 \to x^1 + F(x^0)$ when $F$ obeys
\be \dot F - ({E\0(\pi_1)^2} + {1\0 \b}) \ F(x^0) = f(x^0); \nonumber \ee
or even simpler by observing that due to \re{dQ}) a shift in $f$ does not
change the Hamiltonian (up to boundary terms).
Thus around
points on the space-time manifold with nonvanishing torsion the
geometrical quantities can be always brought into the form
\re{gauge}, \ref{invert} - \ref{lm}) with $f\equiv 0$.

Let us analyse the solutions in the neighbourhood of points with
vanishing torsion, which according to \re{cov}) are also extremal
points of the curvature. Since it is instructive, we will do this
simultanously for both signatures $(-)^M$. Due to the choice of signs
in \re{action}) the field equations \re{cov}) are valid also for $M=1$;
beside the different tangent space metric the difference of the signature
enters only in the definition of the momenta: $\pi_a = (-)^M \b T_a,\ \pi_2 =
(-)^{M+1} \g R$. Differentiating the second equation of \re{cov}) covariantly
and then using the first one, one obtains
\be \pi_{2,a;b} = (-)^M g_{ab} E. \la{ind} \ee
Since at extremal points of $\pi_2$ the covaraint derivative
at the lefthand side of \re{ind}) can be replaced by a normal one,
and since at  a point of vanishing torsion $E=0$ only in the
(Anti-)de-Sitter case of constant curvature,
this equation tells us the following: whereas for Minkowskian signature
extremal points of the curvature are always  saddle points
(cf.\ also \cite{Can})), in the Euclidean case they are true maxima or
minima. This allows us to choose a frame and coordinates
such that
\be \pi_0 = x^0, \qquad \pi_1 = x^1. \la{exg1}  \ee
Note that according to \re{Q}) lines of constant curvature are also
lines of constant $\pi^2 \equiv \pi_0^2 + (-)^M \pi_1^2$, so that
\re{exg1}) captures exactly the different character of
the extremal points of $\pi_2$  for each signature.
As the third gauge condition we require
\be \o_1 =0. \la{exg2} \ee
To see that this gauge is attainable on-shell, one first calculates
the corresponding Faddeev-Popov determinant. One can  verify then that
the determinant does not vanish at the origin since
$E(x^\m=0)\neq 0$; due to the continuity of the solutions this
implies that the determinant does not vanish in a neighborhood
of the considered point $\pi_a=0$.
Finally
one has to check that the resulting solutions satisfy $e\neq 0$,
which is indeed the case here. The
remaining steps  to find the solutions are the same as before:
The inversion of the constraints yields
\be e_1{}^1=0, \qquad e_1{}^0 = {1\0E},  \ee
and $\pi_2$ is determined implicitely through \re{Q})
(with $Q = Q_0 = const$); note that there can exist several branches
of $\pi_2$ (or even none), depending on the value of $Q_0$.
The equations for the Lagrange multipliers give:
\be e_0{}^0 = - {\o_0 x^0 \0 E},
\qquad  e_0{}^1 =  {(-)^{M+1}\o_0 x^1 -1 \0 E},
\qquad \6 \o_0 = {1\0 2\g} e\pi_2. \ee
The last first order differential equation cannot be solved by means of
elementary functions,
since $\pi_2$ cannot be inverted analytically from \re{Q}).
Nevertheless, the existence theorems for such differential equations
guarantee us the $C^\infty$-structure of the resulting solutions
around  points of vanishing torsion.

Although we do not treat all the global aspects of the
classical solutions to \re{action})
within this letter, the above analysis allows an intersting
conclusion about a difference between global Minkowskian and global
Euclidean solutions. For Minkowskian signature the only
maximally extended solutions with nowhere diverging curvature (and
torsion) are the de-Sitter  and Anti-de-Sitter  space \cite{Com}.
For Euclidean signature of \re{action}), however, there exist additional
globally complete solutions with everywhere bounded curvature and torsion.
As one learns from \re{Q}), depending on $Q_0$ there exist 0, 1, 2, or 3
values of $\pi_2$ for which $\pi^2$ vanishes. Let us choose $Q_0$ such that
there exist at least two such values $B_1$ and $B_2$ of $\pi_2$. If one
starts from a point of the space time manifold with  $B_1 < \pi_2 < B_2$,
one can take the fields in  the form of  \re{gauge}, \ref{invert} - \ref{lm}).
This solution extends to a region $\pi_2 = x^0 \in \: ]B_1,B_2[$ since
within this region $\pi^2 \neq 0$. At $\pi_2 = B_i$, however, $\pi^2 =0$
and contrary to the Minkowskian signature this also implies $\pi_a =0$.
According to our considerations following equation \re{ind})
this implies that $\pi_2 = B_1$ is a minimum of the curvature and
$\pi_2 = B_2$ a maximum. Thus in the Euclidean case there exist global
solutions with oscillating curvature and torsion.

This completes the present analysis of the Euclidean solutions to
\re{action}). However, let us  by means of the aquired knowledge
about the solutions of the field equations test our gauge fixing
procedure to find all the classical solutions. From \re{fp}) we
see no reason (as far as $\pi_1 \neq 0$) to not choose e.\ g.
\be \pi_2=const, \qquad \pi_0 = 0, \qquad e_1{}^1 = k(x^\m) \la{fal} \ee
with some function $k$ fulfilling $\dot k \neq 0$ as our gauge fixing
conditions.
{}From the resulting space-time manifolds, on the other hand, we know that
with this gauge choice we either are on a de-Sitter  space or we
are stuck to a {\em line} of constant curvature. The frame work
passes this test: For constant $\pi_2$ the $G_2=0$ constraint as well as
$\6_0 \pi_2 =0$ yield, respectively
\ba &\pi_0 e_1{}^1 - \pi_1 e_1{}^0 =0& \nn
&\pi_0 e_0{}^1 - \pi_1 e_0{}^0 =0.& \nonumber \ea
{}From this we learn that {\em either} $\pi_a=0$, in which case these two
equations are empty and a further analysis, equivalent to the use of
\re{cov}),  leads to the above mentioned (Anti-)de-Sitter
solution, {\em or}
\be \6_0 \quad \propto \quad \6_1, \la{pro} \ee
telling us that the flow of the Hamiltonian indeed does not leave
the line with the initial values. Since \re{pro}) implies $e\equiv0$
of the  resulting solutions one is warned automatically
that something is wrong with the 'gauge' \re{fal}).

Let us turn to the Dirac quantization  of the Euclidean  form of \re{action}).
 For this purpose we choose a momentum
representation for the wave functionals and the operator ordering within
the quantum version of the constraints (4) such that all functional
derivatives have been put to the right; the constraint algebra has no
anomalies then since it still satisfies
\addtocounter{equation}{1}
$$ [G_a, G_2] \, =\,  \varepsilon_{ab} \, G_b \, \delta
\eqno(\arabic{equation}a) $$
$$ [G_a, G_b] \,=\, \varepsilon_{ab} \,
(-\frac{\textstyle 1}{\textstyle \b} \pi_c \, G_c +
\frac{\textstyle 1}{\textstyle 2\gamma}
\, \pi_2 \, G_2) \, \delta .
\eqno(\arabic{equation}b) $$
To find the physical wave functionals, which are defined as the kernel
of the quantum constraints, we can
again have recourse to \cite{Can}, where we obtained the physical wave
functionals of the Minkowskian theory. Making use of \re{pm}), we only
have to substitute $i\hbar$ by $\hbar$ due to the imaginary $\e_{+-}$.
Then we know from there that except for distributional
functionals located at $\pi_a =0$, the kernel of the constraints is given
by the functionals
\be \Psi = \exp(\pm{1\0 \hbar} \oint \ln \pi_\pm d\pi_2) \;
\ti \Psi [Q] ,\qquad \6 Q  \, \ti \Psi[Q]=0.
\la{sol} \ee
The second equation is a restriction on the support of $\ti \Psi[Q]$,
due to which the first equation holds for either of the two signs with
the {\em same} $\ti \Psi[Q]$ --- as far as one has compactified the
$x^1$--direction. In the Dirac quantization it is at this point where the
boundary conditions are to be specified consistently. Since
$\2 \ln({\pi_+\0 \pi_-}) = i \arctan {\pi_1\0 \pi_0}$ the
Euclidean physical wavefunctionals can be rewritten as
\be \Psi = \exp({i\0 \hbar} \oint_{S^1} \Th \6 \pi_2 dx^1) \;
\ti \Psi [Q] ,\qquad \6 Q   \Psi =0, \la{wav} \ee
when using the polar coordinates $(\pi^2, \Th)$ instead of $(\pi_0,\pi_1)$.
Obviously the set of physical wave functionals emerges as
equivalent to the set of ordinary functions $\hat \Psi (Q_0)$, as one would
expect from \re{can}) or \cite{Can}. To find the appropriate inner product,
we require the Dirac observables $Q_0$ and $P_0$ to become hermitian.
Substituting $e_1{}^a$ by $i\hbar(\d /\d \pi_a)$ in \re{P0}) and
applying this operator to \re{sol}) one finds that $P_0$ acts as
$-i\hbar (d/dQ_0)$ on $\hat \Psi(Q_0)$. Therefore the measure within
the inner product has to be the Lebesgue measure,
\be \< \Psi, \Phi \> = \int dQ_0 \hat \Psi(Q_0)^* \hat \Phi(Q_0), \la{inn} \ee
so that we end up with an $\CL^2(R)$ as in the Minkowskian case. Since
to treat the 'issue of time' within a Euclidean theory seems somewhat
artificial,
we will skip this point here. (The reader interested in this topic ---
there is no meaningful Schroedinger equation since the Hamiltonian
vanishes on \re{wav}) ---  shall be refered to the Minkowskian
counterpart in \cite{Can}).

As a next step one should extend the present considerations
to the case of a  string action coupled to \re{action}).
For the Minkowskian signature this has been done quite recently by
Katanaev \cite{String}.

\begin{appendix}
\section{Appendix}

In general, gauge conditions on metric and connection, which can be
realized by diffeomorphisms and local Lorentz transformations
on a space with Minkowski type metric are not available in a
Euclidean space. Complexifying the tangent bundle over the space time
manifold, it may, although, be possible to apply such gauges to the
Euclidean case.

The method is well established for the gauge condition
$g_{00}=g_{11}=0$. Given any metric there will
always exist two lightlike independent vector fields
in the complexified tangent bundle.
We may then find two (in the Euclidean case complex)
functions $z_0$, $z_1$ such that each of them is constant in one of the
lightlike directions and the exterior derivatives of these functions form a
basis in the complexified cotangent bundle. In this basis the metric
obviously has the form $g=f\,dz_0\otimes dz_1$
with a complex coefficient function f.
As g is real we have $g=\bar g=\bar f\,d\bar z_0d\bar z_1$.
With the Ansatz $d\bar z_0=adz_0+bdz_1$, $d\bar z_1=cdz_0+edz_1$
on deduces easily that either $b=c=0$ or $a=e=0$.
The first case corresponds to the
Minkowski signagture. In the second case we have
$dz_1=(1/b)d\bar z_0$. From $d^2=0$ we conclude that
b depends on $\bar z_0$ only.
After calculating $b$ from the conditions $g=\bar g$
and ${\bar{\bar z\hskip 0.1 em}\hskip -0.1 em}_i=z_i$
we can express $z_1$ as a function of $\bar z_0$ and
then return to real coordinates
$x_0=z_0+\bar z_0$ and $x_1=i(z_0-\bar z_0)$.

The aim of this appendix is, to apply this method to the
local solutions of the equations of
motion of the Lagrangian \re{action}) which
were calculated in \cite{Kum} in a gauge characterized by the conditions
$e_0{}^+=1$, $e_0{}^-=0$, $\omega_0=0$ (light cone gauge).
It is well known that this gauge is
always obtainable for a Minkowski-type metric.

In a Euclidean space,
however, we need complex coordinates as well as a complex
local rotation to achieve this gauge.
With some additional gauge fixing we then find the general
solution in the neighborhood of a point, where
$\pi^2\neq 0$ to be determined by a real constant $Q_0$: \\
     {$\vcenter{\vbox {  $$  \displaylines {
       \pi_{\omega} =   x^0, \qquad \pi_+ = i  \qquad
       \pi_- =  {-i\0 2}W(x^0) \cr
 \omega_1 =- e_1{}^- \, \dot\pi_-, \qquad e_1{}^+ = -ie_1{}^- \, \pi_- \cr
    e_1{}^- =  \exp(x^0/\beta) \cr}
      $$ }} \hskip -\hsize \vcenter{\vbox { $$ \eqno(A1)$$}}$}
The quantity $W(x^0)$ was defined in (15).

As $\pi_2$ is a function on the space time manifold, we have
$\bar\pi_2=\pi_2$ and thus $x^0$ turns out to be real.
This also guarantees $\pi^2$ to be real.
With the ansatz
   $$d\bar z^1=adx^0+bdz^1 \eqno(A2)$$
in $g=\bar g$, where the metric g
is given by the expression
   $$ g= 2\exp(x^0/\beta)dx^0dz^1-
   2i\exp(2x^0/\beta)\pi_- dz^1dz^1, \eqno(A3)$$
we obtain a set of equations for a and b:
$$  Aa^2+a=0 \quad 2Aab+b-1=0 \quad Ab^2-A=0  \eqno(A4)$$
where $A \equiv -i\exp(x^0/\beta)\pi_-$ .
They
are simultaneously solved by
$$a=-{1\over A}, \quad b=-1  \eqno(A5)$$
(A second solution $a=0$, $b=1$ corresponds to the Minkowski space).
As a check of consitency one may verify that
$d^2 \bar z^1 =0$ and $d\bar{\bar z\hskip 0.1 em}^1=dz^1$.
We may now introduce the real coordinate $x^1=i(\bar z^1-z^1)$ and express
$z^1$ in terms of $x^0$ and $x^1$:
$$z^1={1\over2}(ix^1 - \int{1\over A} dx^0) \eqno(A6)$$
Note that the solution (A1) does not depend on $z^1$. Thus we
need not solve the integral in (A6). To calculate the Euclidean
solutions it is sufficient to know $dz^1$ as a linear combination
of the $dx^i$.

But still the components of the torsion
$\pi_0={1\over \sqrt 2}(\pi_++\pi_-)$ and
$\pi_1={-i\over \sqrt2}(\pi_+-\pi_-)$
as well as the connection $\omega$ are not real. This is not surprising,
as we had to apply a complex local rotation to achieve the light
cone gauge. It is easy to verify that $\pi_0$, $\pi_1$, and $\omega $ become
real by the transformation
$$ \pi_+\to\alpha\pi_+ \quad \pi_-\to{1\over\alpha}\pi_-, \quad
   \omega\to\omega+id\ln \alpha\; ; \quad      \alpha=\sqrt{{W\0 2}}
   \eqno(A7)$$
The construction of the Euclidean solution is thus complete:
$$ \pi_2  =x^0 \qquad \pi_0 = 0 \qquad \pi_1 = \sqrt{W} $$
$$ \omega _0=0 \qquad \omega _1=- \4 \dot W\, \exp(x^0/\beta)
   \eqno(A8)$$
$$ {e_0}^0 = \sqrt{1\over W} \qquad {e_1}^0={e_0}^1=0 \qquad
      {e_1}^1= \2 \sqrt{W}\, \exp(x^0/\beta) $$

This relult is transformed into the conformal gauge
($g_{01}=0$, $g_{00}=g_{11})$ by the coordinate transformation\\
         $\vcenter{\vbox{
    $$\displaylines{
      \tilde x^0 = \int {2 \over W(x^0)} \exp(x^0/\beta) dx^0
      \qquad \tilde x_1 = x_1   \cr
      \Rightarrow \qquad g= \4 W(h(\tilde x^0))\exp(2h(\ti x^0)/\beta)
       (d\tilde x^0 d\tilde x^0 + d\tilde x^1 d\tilde x^1) \cr
        \omega =- \4  W^\prime(h(\ti x^0)) \, \exp(h(\ti x^0)/\beta) dx^1
        \cr    }$$}}
     \hskip -\hsize \vcenter{\vbox{$$\eqno (A9)$$}}$
where $h(\ti x^0) = \pi_2(\ti x^0)$ is the inverse function of
$\tilde x^0 (x^0)$ and solves Katanaev's differential equation
\cite{Kat} $\dot h = \2 W(h)\exp (h/\beta)$.  (A9) agrees with
the result obtained by a Wick rotation of the solution given in
\cite{Kat}.

\end{appendix}

\end{document}